\null

\input epsf
%*******************************************************************
%FONTS

\magnification 1200
\font\bigbf= cmbx10 scaled \magstep1

%*******************************************************************
%FOOTNOTES - insert command \eightpoint to reduce

\newskip\ttglue

\font\eightrm=cmr8
\font\eighti=cmmi8
\font\eightsy=cmsy8
\font\eightbf=cmbx8
\font\eighttt=cmtt8
\font\eightsl=cmsl8
\font\eightit=cmti8
\font\sixrm=cmr6
\font\sixbf=cmbx6
\font\sixi=cmmi6
\font\sixsy=cmsy6

\def \eightpoint{\def\rm{\fam0\eightrm}% switch to 8-point type
\textfont0=\eightrm \scriptfont0=\sixrm \scriptscriptfont0=\fiverm
\textfont1=\eighti \scriptfont1=\sixi \scriptscriptfont1=\fivei
\textfont2=\eightsy \scriptfont2=\sixsy \scriptscriptfont2=\fivesy
\textfont3=\tenex \scriptfont3=\tenex \scriptscriptfont3=\tenex
\textfont\itfam=\eightit \def\it{\fam\itfam\eightit}%
\textfont\slfam=\eightsl \def\sl{\fam\slfam\eightsl}%
\textfont\ttfam=\eighttt \def\tt{\fam\ttfam\eighttt}%
\textfont\bffam=\eightbf \scriptfont\bffam=\sixbf
\scriptscriptfont\bffam=\fivebf \def\bf{\fam\bffam\eightbf}% \tt
\ttglue=.5em plus.25em minus.15em
\setbox\strutbox=\hbox{\vrule height7pt depth2pt width0pt}%
\normalbaselineskip=9pt
\let\sc=\sixrm \let\big=\eightbig \normalbaselines\rm }
%******************************************************************* 
%GREEK LETTERS
\def\a{\alpha}
\def\b{\beta}
\def\c{\gamma}
\def\d{\delta}
\def\e{\epsilon}

\def\l{\lambda}
\def\m{\mu}
\def\n{\nu}
\def\o{\theta}
\def\p{\pi}
\def\r{\rho}
\def\s{\sigma}
\def\t{\tau}

\def\w{\omega}

\def\C{\Gamma}
\def\D{\Delta}
\def\L{\Lambda}

\def\S{\Sigma}

%***************************************************************** 
%OTHER MACROS
\def\pl{\partial}
\def\ra{\rightarrow}

\def\DD{{\cal D}}

\def\OO{{\cal O}}
\def\BB{{\cal B}}

\def\GV{{\rm GeV}}
\def\MV{{\rm MeV}}

\def\el{{\rm e}}

\def\SIMQ{\mathrel{\mathop \sim_{Q^2\rightarrow\infty}}} 
\def\SIMEQz{\mathrel{\mathop \simeq_{z \sim 1}}}

\def\ul{\underline}

%***************************************************************** 
%TITLE PAGE

{\nopagenumbers

\line{\hfil SWAT 98/212 }
\line{\hfil CERN--TH/98--398 }
\vskip0.7cm

%\vbox to 1.8cm { }

\centerline{\bigbf The Proton-Spin Crisis: another ABJ anomaly?
%${}^*$
}
\vskip0.7cm
\centerline{\bf G.M. Shore}
\vskip0.5cm
\centerline{\it Department of Physics, University of Wales Swansea} 
\centerline{\it Singleton Park, Swansea, SA2 8PP, U.K. }
\vskip0.2cm
\centerline{and} 
\vskip0.2cm
\centerline{\it Theory Division, CERN}
\centerline{\it CH 1211 Geneva 23, Switzerland} 
\vskip0.7cm
\centerline{\bf Contents:}
\vskip0.3cm

\line{\hskip2.5cm 1.~Introduction \hfil}
\vskip0.1cm
\line{\hskip2.5cm 2.~The First Moment Sum Rule for $g_1^p$ \hfil}
\vskip0.1cm
\line{\hskip2.5cm 3.~The Parton Model and the `Proton Spin' \hfil}
\vskip0.1cm
\line{\hskip2.5cm 4.~The CPV Method and Topological Charge Screening \hfil}
\vskip0.1cm
\line{\hskip2.5cm 5.~Experiment \hfil}
\vskip0.1cm
\line{\hskip2.5cm 6.~Semi-Inclusive Polarised DIS \hfil}

\vskip0.7cm
\noindent {\bf 1.~~Introduction}
\vskip0.3cm
For a decade, the `proton spin' problem -- the anomalous suppression
observed in the flavour singlet component of the first moment of the 
polarised proton structure function $g_1^p(x;Q^2)$ -- has puzzled and
intrigued theorists and experimentalists alike. The consequent research
effort has indeed been impressive:~to date, the original EMC paper[1] 
alone has nearly one thousand citations.

In this lecture, we first give a brief review of the `proton spin'
problem from the standard viewpoint of the parton model. We explain
why the 1988 observation by the EMC of a violation of the Ellis-Jaffe
sum rule[2] for $g_1^p$ was initially mis-interpreted in terms of
quark spins and how the problem is resolved in the context of the
full QCD parton model[3].

\vskip0.5cm
\noindent ${}^*$ ~{\it Lecture presented at the International School
of Subnuclear Physics, `From the Planck Length to the Hubble Radius',
Erice, 1998.}

\vskip0.3cm

\line{SWAT 98/212 \hfil}
\line{CERN--TH/98--398 \hfil}
\line{December 1998 \hfil}

\vfill\eject }

\pageno=2

We then describe an alternative, complementary approach to the description 
of deep inelastic scattering (DIS), the `CPV' method[4-6], which allows the 
problem to be viewed in a new light. From this perspective, the
Ellis-Jaffe sum rule is simply seen to be equivalent to the OZI approximation 
for the forward proton matrix element of the flavour singlet axial current.
The `proton spin' problem is therefore one of understanding the origin of the
OZI violation observed in this channel[7]. As such, it is one more addition to 
the collection of `$U_A(1)$ problems' in QCD -- phenomena whose
interpretation depends on the presence of the ABJ axial anomaly[8] and 
the implicit relation with gluon topology (see e.g.~ref.[9]). 
As we shall show, the 
observed suppression in the first moment of $g_1^p$ is due to
{\it topological charge screening} by the QCD vacuum itself, and a quantitative
resolution in terms of an anomalous suppression of the slope of the
gluon topological susceptibility is proposed.

An immediate consequence of this explanation is that the suppression
in $g_1^p$ is in fact a {\it target independent} phenomenon, which would
in principle be true for polarised DIS on any hadronic target. 
Not only is the `proton spin' problem nothing to do with spin -- it is
not even a special property of the proton! To test this idea, we have 
proposed[10,11] a set of semi-inclusive polarised DIS experiments, which 
could be performed at e.g.~polarised HERA, and which would provide
independent confirmation of the mechanism of topological charge screening
by the QCD vacuum.

\vskip1cm

\noindent {\bf 2.~~The First Moment Sum Rule for $g_1^p$}
\vskip0.5cm
The structure function $g_1^p$ is measured in polarised DIS experiments
through the inclusive processes $\m p \ra \m X$ (EMC, SMC) or
$e p \ra e X$ (SLAC, HERMES). See Fig.~1. 
\vskip0.2cm
\centerline{
{\epsfxsize=3.5cm\epsfbox{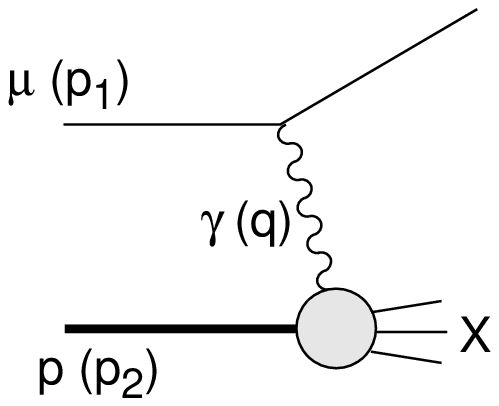}}}
\noindent{\eightrm Fig.1~~Inclusive polarised DIS scattering.}
\vskip0.2cm
The polarisation asymmetry of the cross section is expressed as
$$
x{d\D\s\over dx dy}~=~{Y_P\over2}{16\pi^2\a^2\over s} g_1^p(x,Q^2)~~+~~
O\Bigl({M^2 x^2\over Q^2}\Bigr)
\eqno(2.1)
$$
where the omitted terms include the second polarised structure function 
$g_2^p$. The notation is standard: $Q^2 = -q^2$ and $x = {Q^2\over2p_2.q}$
are the Bjorken variables, $y = {Q^2\over xs}$ and $Y_p = {2-y\over y}$.

According to standard theory, $g_1^p$ is determined by the proton
matrix element of two electromagnetic currents carrying a large spacelike
momentum. The sum rule for the first moment (w.r.t. Bjorken $x$)
of $g_1^p$ is derived using the twist 2, spin 1 terms in the OPE for
the currents:
$$
J^\r(q) J^\s(-q) ~\SIMQ~
2 \e^{\r\s\n\m} {q_\n\over Q^2} ~\Bigl[C_1^{NS}(\a_s) 
\Bigl(A_{\m }^3 + {1\over\sqrt3} A_{\m }^8\Bigr)
+ {2\over3} C_1^{S}(\a_s) A_{\m }^0 \Bigr]
\eqno(2.2)
$$
where $C_1^{NS}$ and $C_1^S$ are the Wilson coefficients and $A_\m^a$
($a=0,3,8$) are the renormalised $SU(3)$ flavour axial currents.
The sum rule is therefore:
$$
\C^p_1(Q^2) \equiv
\int_0^1 dx~ g_1^p(x,Q^2) 
= {1\over12} C_1^{NS} \Bigl( a^3
+ {1\over3} a^8 \Bigr) + {1\over9} C_1^{\rm S} a^0(Q^2)  
\eqno(2.3)
$$
where the axial charges $a^3$, $a^8$ and $a^0(Q^2)$ are defined as 
the form factors in the forward proton matrix elements of the 
axial current, i.e.
$$
\langle p, s|A_{\m }^3|p, s\rangle = s_\m {1\over2} a^3 ~~~~~~~~
\langle p, s|A_{\m }^8|p, s\rangle = s_\m {1\over{2\sqrt3}} a^8 ~~~~~~~~
\langle p, s|A_{\m }^0|p, s\rangle = s_\m a^0(Q^2)
\eqno(2.4)
$$
Here, $p_\m$ and $s_\m = \bar u \c_\m\c_5 u$ are the momentum and polarisation 
vector of the proton respectively. 

It is important in what follows to be precise about the renormalisation group
properties of the operators and matrix elements which occur. The flavour
non-singlet axial currents are not renormalised, but because of the 
$U_A(1)$ anomaly the flavour singlet current $A_\m^0$ is not conserved and
therefore can, and does, have a non-trivial renormalisation. Defining the bare
operators by
$A_{\m B}^a = \sum \bar q \c_\m \c_5 T^a q$
and $Q_B = {\a_s\over{8\p}} \e^{\m\n\r\s}{\rm tr} G_{\m\n} G_{\r\s}$, 
(where $T^{a\neq0}$ are $SU(n_f)$ flavour generators, $T^0 = {\bf 1}$,
and $d$-symbols are defined by $\{T^a,T^b\} = d_{abc}T^c$), the renormalised
operators $A_\m$, $Q$, are given by[12] 
$$
A_\m^{a\neq0} = A_{\m B}^{a\neq0}~~~~~~~~
A_\m^0 = Z A_{\m B}^0 ~~~~~~~~
Q = Q_B - {1\over2n_f}(1-Z) \pl^\m A_{\m B}^0 
\eqno(2.5)
$$
where $Z$ is a divergent renormalisation constant, the associated
anomalous dimension being denoted by $\c$. 
$Q$ is the gluon topological charge density -- for field configurations
which tend to a pure gauge at infinity, it satisfies
$$
\int d^4 x ~Q = n \in {\bf Z}
\eqno(2.6)
$$
where $n$ is the topological winding number, or `instanton' number.

The link between quark dynamics in the flavour singlet $U_A(1)$ channel and
gluon topology is provided by the famous $U_A(1)$ axial (or ABJ)
anomaly\footnote{${}^{(1)}$}{\eightpoint
\noindent In this lecture, we will quote results only in the chiral limit
of QCD, i.e.~with massless quarks. The inclusion of quark masses, which
produces only a small change in our final predictions, is described
in detail in ref.[13].}:
$$
\pl^\m A_\m^0 - 2n_f Q \sim 0
\eqno(2.7)
$$
where the symbol $\sim$ denotes weak operator equivalence.
Notice that with the definitions (2.5), this condition is the same expressed 
in terms of either the bare or renormalised operators. 

A more complete formulation of the anomaly is given by introducing the
generating functional $W[S_\m^a, \o, S_5^a, S^a]$ of connected
Green functions (correlation functions) of the axial currents
and the pseudoscalar operators $Q$ and $\phi_5^a$, where
$\phi_{5B}^a = \sum \bar q \c_5 T^a q$ (together with the
corresponding scalar $\phi^a$). Here, $S_\m^a, \o, S_5^a, S^a$ 
are the sources for $A_\m^a, Q, \phi_5^a, \phi^a$ respectively.
The (anomalous) Ward identities are then expressed as 
$$
\pl_\m {\d W\over\d S_\m^a} - 2n_f \d_{a0}{\d W\over\d\o}
+ d_{adc} S^d {\d W \over \d S_5^c}
- d_{adc} S_5^d {\d W\over\d S^c}
= 0
\eqno(2.8)
$$
where the final two terms account for the chiral variations of the
operators $\phi_5^a$, $\phi^a$. (For a more extensive description
of this formalism, see refs.[5,9].)
Ward identities for 2-point Green functions (composite operator
propagators) are then derived by taking functional derivatives w.r.t.~the
relevant sources. For example,
$$
\pl_\m W_{S_\m^0 \o} - 2n_f  W_{\o \o} = 0 
\eqno(2.9)
$$
where for simplicity we adopt a notation where functional derivatives
are denoted by subscripts and integrals over spacetime are assumed
as appropriate.
Since there is no massless $U_A(1)$ Goldstone boson in the physical
spectrum (resolution of the $U_A(1)$ problem), the first term
vanishes at zero momentum, leaving\footnote{${}^{(2)}$}{\eightpoint
\noindent Here, and elsewhere throughout the text, I have
included a discussion of points omitted during the lecture itself
which were subsequently raised in the discussion sessions.}\footnote{
${}^{(3)}$}{\eightpoint
\noindent In fact, the same conclusion would hold even if there were
a physical massless boson coupling derivatively to the $U_A(1)$
current since, in momentum space, the residue of the pole at $k^2=0$
in the first term in eq.(2.9) due to this boson would itself be
$O(k^4)$.}
$$
W_{\o \o}\big|_{k=0} = 0
\eqno(2.10)
$$

The renormalisation group equations (RGEs) implied by eq.(2.5) can
also be conveniently written in a functional form. The fundamental RGE
for the generating functional $W$ is[5,9]
$$
\DD W = \c\Bigl(S_\m^0 - {1\over2n_f}\pl_\m \theta\Bigr)W_{S_\m^0}
+ \c_\phi\Bigl(S_5^a W_{S_5^a} + S^a W_{S^a}\Bigr) + \ldots
\eqno(2.11)
$$
where $\DD = \Bigl(\m{\pl\over\pl\m} + \b{\pl\over\pl g}\Bigr)
\Big|_{V,\theta,S_5,S}$ and $\c_\phi$ is the anomalous dimension
corresponding to the pseudoscalar (or scalar) composite field 
renormalisation. The notation $+\ldots$ refers to the additional terms 
which are required to produce the contact terms in the 
RGEs for Green functions involving more than one composite operator.
These vanish at zero momentum and need not be considered here.

The RGEs for 2-point Green functions
follow immediately by functional differentiation of eq.(2.11).
For example, differentiating twice w.r.t.~$\o$, and using the Ward 
identity (2.9), we find
$$
\DD W_{\o\o} = 2\c W_{\o\o} + \ldots
\eqno(2.12)
$$
We are focusing on results for $W_{\o\o}$ here because it is a key
correlation function in QCD which will play a central role in our
analysis of the `proton spin' problem. It is called the 
{\it topological susceptibility}, and is usually denoted by $\chi$.
In conventional notation,
$$ 
\chi(k^2) \equiv W_{\o\o}(k^2) = 
i\int d^4x~e^{ikx}\langle 0|T~Q(x)~Q(0)|0\rangle
\eqno(2.13)
$$

Returning to the structure function sum rule (2.3), we can now deduce 
the RG behaviour of the axial charges. From either (2.5) or (2.11), we
see immediately that
$$
{d\over dt} a^{3,8} = 0 ~~~~~~~~~~~
{d\over dt} a^0(Q^2) = \c a^0(Q^2)
\eqno(2.14)
$$
where $t = \ln Q^2/\L^2$. The singlet axial charge is therefore
scale dependent. This is crucial in understanding the `proton spin' problem
in QCD.

The flavour non-singlet axial charges $a^3$ and $a^8$ are known in terms
of the $F$ and $D$ constants found from neutron and hyperon beta decay:
$$
a^3 = F+D ~~~~~~~~~~
a^8 = 3F-D
\eqno(2.15)
$$
The interest of the sum rule therefore centres on the flavour singlet axial
charge $a^0(Q^2)$. In the absence of an alternative experimental 
determination of $a^0(Q^2)$, the simplest ansatz is to assume that it obeys
the OZI rule, i.e. $a^0(Q^2) = a^8$. 
In precise terms, the OZI limit of QCD is defined[14] as the truncation of 
full QCD in which non-planar and quark-loop diagrams are retained, but 
diagrams in which the external currents are attached to distinct quark 
loops, so that there are purely gluonic intermediate states, are omitted. 
(This last fact makes the connection with the familiar phenomenological 
form of the OZI, or Zweig, rule.) This is a more
accurate approximation to full QCD than either the leading large 
$1/N_c$ limit, the quenched approximation (small $n_f$ at fixed $N_c$) 
or the leading topological expansion ($N_c\ra\infty$ at fixed $n_f/N_c$).
In the OZI limit, the $U_A(1)$ anomaly is absent, there is no meson-glueball
mixing, and there is an extra $U_A(1)$ Goldstone boson. 
\vskip0.2cm
\centerline{
{\epsfxsize=9cm\epsfbox{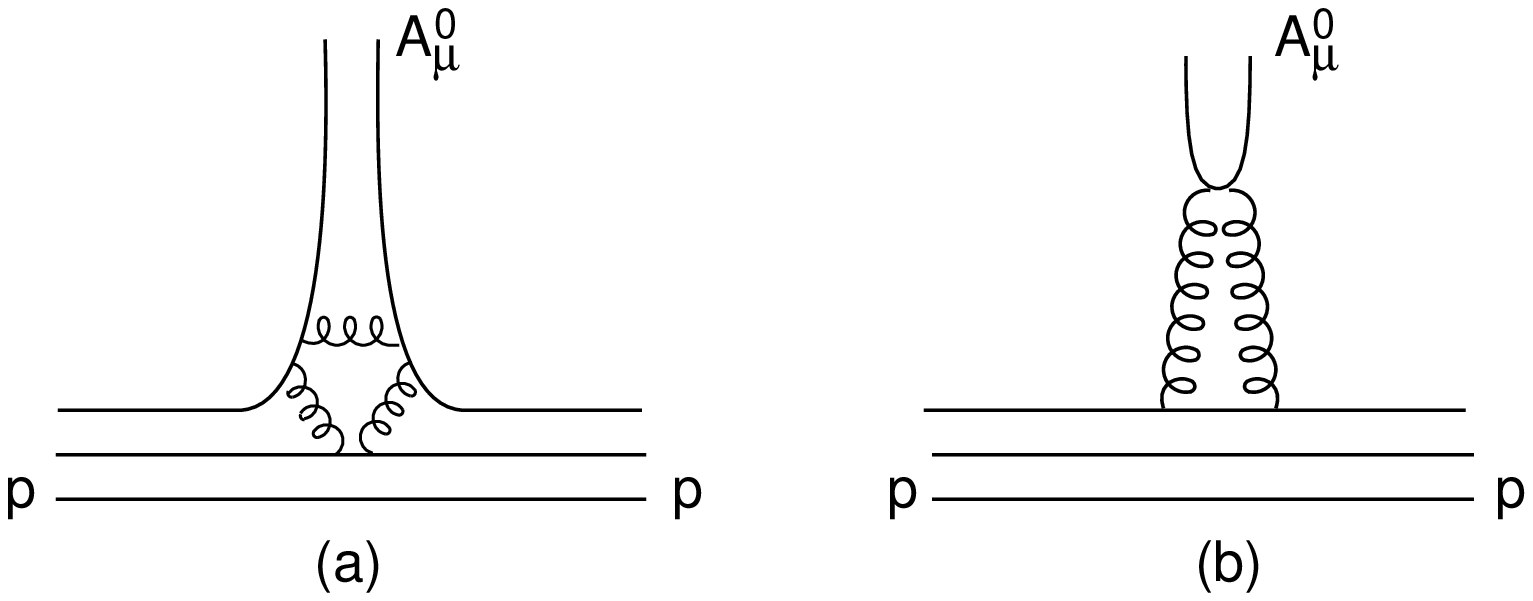}}}
\noindent{\eightrm Fig.2~~`Connected' (a) and `disconnected' (b) contributions
to the matrix element $\langle p|A_\m^0|p\rangle$. Diagrams of type
(b) are suppressed in the OZI limit.}
\vskip0.2cm
Applied to the axial charges, the OZI rule states that of the two valence
quark diagrams shown in Fig.~2 describing the coupling of the axial current 
to the proton, the diagram in Fig.~2(b) is suppressed. Since this is the
only way the $s$-quark component of either $A_\m^0$ or $A_\m^8$ can
couple to the proton ($uud$), the OZI prediction $a^0 = a^8$
follows immediately.

Inserted into the sum rule (2.3), the OZI approximation $a^0(Q^2) = a^8$
gives a theoretical prediction for $\C_1^p$ which is known as the
Ellis-Jaffe sum rule. This is now known to be violated, with
$a^0(Q^2)$ strongly suppressed (by a factor of around 0.5) relative to
$a^8$. This experimental result, which was first discovered by the
EMC collaboration and confirmed and improved in the subsequent decade
of experimental work, is what has come to be known as the `proton spin'
problem (or even crisis!).

In fact, it is not at all surprising that the OZI rule should fail in this
case[7,4,5]. The first clue is the anomaly-induced scale dependence of 
$a^0(Q^2)$. If the OZI rule were to hold, at what scale should it be applied?
Moreover, it is known that the pseudovector and pseudoscalar channels are 
linked through the Goldberger-Treiman relations. Since large anomaly-induced 
OZI violations are known to be present in the pseudoscalar channel 
($U_A(1)$ problem, $\eta'$ mass, etc.) it is natural to find them also 
for $a^0(Q^2)$ in the pseudovector channel.
It is also immediately clear from its scale-dependence that $a^0(Q^2)$ cannot
really measure spin.

While this immediately resolves the `proton spin' {\it problem}, clearly
we want to understand the origin of the suppression in $a^0(Q^2)$ much more
deeply. In the next two sections, we describe two complementary approaches
to this question -- the conventional QCD parton model and the
CPV method developed in refs.[4-6].

\vskip1cm

\noindent{\bf 3.~~The Parton Model and the `Proton Spin'}
\vskip0.5cm
In the simplest form of the parton model, the proton structure for large 
$Q^2$ DIS is described by parton distributions corresponding to free valence
quarks only. The polarised structure function is given by
$$
g_1^p(x) = {1\over2} \sum_{i=1}^{n_f} ~e_i^2 ~\D q_i(x)
\eqno(3.1)
$$
where $\D q_i(x) = q_i^+(x) + \bar q_i^+(x) - q_i^-(x) - \bar q_i^-(x)$
is defined as the difference of the distributions of quarks (and antiquarks)
with helicities parallel and antiparallel to the nucleon spin. 
It is convenient to work with the flavour non-singlet 
and singlet combinations:
$$
\D q^{NS}(x) = \sum_{i=1}^{n_f} \Bigl({e_i^2\over\langle e^2\rangle} 
- 1\Bigr)~ \D q_i(x) ~~~~~~~~
\D q^S(x) = \sum_{i=1}^{n_f} ~\D q_i(x)
\eqno(3.2)
$$
In this model, the first moment of the singlet quark distribution
$\D q^S = \int_0^1 dx~\D q^S(x)$ can be identified as the sum of
the helicities of the quarks. Interpreting the structure function
data {\it in this model} then leads to the conclusion that the quarks carry 
only a small fraction of the spin of the proton -- the `proton spin' problem.
There is indeed a real contradiction between the experimental data
and the {\it free valence quark} parton model.

However, this simple model leaves out many important features of QCD, the most 
important being  gluons, RG scale dependence and the chiral $U_A(1)$ anomaly. 
When these effects are included, in the QCD parton model, the naive 
identification of $\D q^S$ with spin no longer holds and the experimental 
results for $g_1^p$ are readily accommodated.\footnote{${}^{(4)}$}{\eightpoint
\noindent Of course there is a separate angular momentum sum rule for the 
proton. The details are given in e.g.~ref.[15]. 
The proton spin is given by the matrix element of the angular momentum 
operator $\ul{J}$:
$$
J_i = {1\over2}\e_{ijk}\int d^3\ul{x} M^{0jk}~, ~~~~~~~~
M^{\l\m\n} = x^\m T^{\n\l} - x^\n T^{\m\l}
$$
where $T_{\m\n}$ is the energy-momentum tensor. This admits
(up to equation of motion terms) a gauge-invariant
decomposition into three terms which look like quark spin,
quark orbital and total gluon angular momenta:
$$
\ul{J} = \int d^3\ul{x}~\Bigl[\sum q^{\dagger} \ul{\c} \c_5 q + 
\sum q^{\dagger}(\ul{x}\times i\ul{D})q + 
\ul{x}\times(\ul{E}\times \ul{B})\Bigr]
$$
We recognise here the operator $A_{\m B}^0 = \sum \bar q \c_\m \c_5 q$ which,
{\it for free fields}, can be identified as the quark helicity operator.
However, the composite operators in this expression must be renormalised
and their scale dependence identified, and the above identifications then
become more problematic. While it is interesting to pursue both the 
theory and experimental consequences of this sum rule[15], it is
obvious that the axial charge $a^0(Q^2)$, given by the matrix element of 
the renormalised current $A_{\m}^0$, is {\it not} measuring the spin of 
the proton.}
\vskip0.2cm
\centerline{
{\epsfxsize=3.5cm\epsfbox{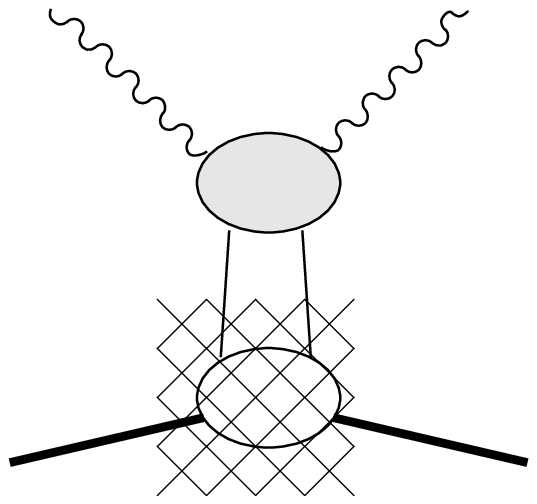}}}
\noindent{\eightrm Fig.3~~QCD parton model interpretation of DIS.
The single lines are partons, which may be quarks or gluons.}
\vskip0.2cm
The QCD parton model picture of DIS is shown in Fig.~3.
The polarised structure function is written in terms of both quark and
gluon distributions[3] as follows:
$$
g_1^p(x,Q^2) = {1\over9}\int_x^1 {dy\over y}\Bigl[
C^{NS}\Bigl({x\over y}\Bigr)\D q^{NS}(y,t) 
+ C^{S}\Bigl({x\over y}\Bigr)\D q^S(y,t) +
C^{g}\Bigl({x\over y}\Bigr)\D g(y,t) \Bigr]
\eqno(3.3)
$$
where $C^S, C^g$ and $C^{NS}$ are perturbatively calculable functions
related to the Wilson coefficients in sect.~2 and the quark and gluon
distributions have {\it a priori} a $t=\ln Q^2/\L^2$ dependence.

The RG evolution (DGLAP) equations for these polarised distributions
are:
$$
{d\over dt}\D q^{NS}(x,t) = {\a_s\over2\pi} \int_x^1 {dy\over y}
P_{qq}^{NS}\Bigl({x\over y}\Bigr) \D q^{NS}(y,t)
\eqno(3.4)
$$
and, 
$$
{d\over dt} \left(\matrix{\D q^S(x,t) \cr \D g(x,t)\cr }\right) =
{\a_s\over2\pi} \int_x^1 {dy\over y}
\left(\matrix{P_{qq}^{S}({x\over y}) & P_{qg}({x\over y}) \cr 
P_{gq}({x\over y}) & P_{gg}({x\over y}) \cr }\right)
\left(\matrix{\D q^S(y,t) \cr \D g(y,t)\cr }\right)
\eqno(3.5)
$$
showing the mixing between the singlet quark and the gluon distributions.
The splitting functions $P$ are also calculable in perturbative QCD,
their moments being related to the anomalous dimensions of the series of
increasing spin operators appearing in the OPE (2.2).

In this language, the first moment sum rule for $g_1^p$ reads:
$$
\C_1^p(Q^2) = {1\over9}\Bigl[C_1^{NS}\D q^{NS}
+ C_1^S \D q^S + C_1^g \D g \Bigr]
\eqno(3.6)
$$
where $\D q^{NS}$, $\D q^S$ and $\D g$ are the first moments of the above 
distributions.
Comparing with eq.(2.3), we see that the axial charge $a^0(Q^2)$ is identified 
with a linear combination of the first moments of the singlet quark
and gluon distributions. It is often, though not always, the case that
the moments of parton distributions can be identified in one-to-one
correspondence with the matrix elements of local operators. The polarised
first moments are special in that two parton distributions correspond
to the same local operator. 

The RG equations for the first moments of the parton distributions follow
immediately from eqs.(3.4,3.5) and depend on the matrix of anomalous dimensions
for the lowest spin, twist 2 operators. This introduces a renormalisation
scheme ambiguity. The issue of scheme dependence has been studied
thoroughly by Ball, Forte and Ridolfi[16] and an excellent summary can 
be found in ref.[17]. It is shown there that it is possible to 
choose a scheme known as the Adler-Bardeen or AB scheme (strictly,
a class of schemes[16]) for which the parton
distributions satisfy the following RG equations:
$$\eqalignno{
&{d\over dt} \D q^{NS} = 0 ~~~~~~~~~~
{d\over dt} \D q^S = 0 \cr
&{} \cr
&{d\over dt} {\a_s\over2\pi}\D g(t) = \c \Bigl({\a_s\over2\pi}\D g(t) 
-{1\over n_f} \D q^S\Bigr)
&(3.7) \cr }
$$
with the implication $C_1^g = -n_f{\a_s\over2\p} C_1^S$.
It is then possible to make the following identifications with the axial
charges:
$$\eqalignno{
a^3 &= \D u - \D d \cr
%{}&{}\cr
a^8 &= \D u + \D d - 2 \D s \cr
%{}&{}\cr
a^0(Q^2) &= \D q^S - n_f {\a_s\over2\pi} \D g(Q^2)
&(3.8) \cr }
$$
where $\D u = \int_0^1 dx \bigl(\D u(x,t) + \D\bar{u}(x,t)\bigr)$ etc.
Notice that in the AB scheme, the singlet quark distribution $\D q^S =
\D u + \D d + \D s$
(which is often written as $\D \S$) is scale independent. All the scale
dependence of the axial charge $a^0(Q^2)$ is assigned to the gluon distribution
$\D g(Q^2)$. We emphasise that eq.(3.7) is true only in the AB
renormalisation scheme and that it is only
in this scheme that the identifications (3.8) hold.

This was the identification originally introduced for the first moments
by Altarelli and Ross[3], and resolves the `proton spin' problem in the
context of the QCD parton model. 
In this picture, the Ellis-Jaffe sum rule follows from the assumption
that in the proton both $\D s$ and $\D g(Q^2)$ are zero.
This is the natural assumption in the context of the free valence quark
model. It is equivalent to the naive OZI approximation $a^0(Q^2) = a^8$ 
described above. However, in the full QCD parton model, there is no
reason why $\D g(Q^2)$, or even $\D s$, need be zero in the proton.
Moreover, given the RG scale dependence of $a^0(Q^2)$,
this assumption is {\it in contradiction with} QCD where the anomaly requires
$a^0(Q^2)$ to scale with the anomalous dimension $\c$.

Since neither $\D \S$ nor $\D g(Q^2)$ are currently measurable in other
processes, the parton model is unable to make a quantitative prediction
for the first moment $\C_1^p$. While the model can accomodate the
observed suppression, it cannot yet predict it.

An interesting conjecture, proposed in the original paper of Altarelli 
and Ross[3], is that the observed suppression in $a^0(Q^2)$ is
due overwhelmingly to the gluon distribution $\D g(Q^2)$. If so, the
strange quark distribution $\D s \simeq 0$ in the proton and so
$\D\S \simeq a^8$. Although by no means a necessary consequence of QCD,
this is entirely plausible because it is the 
anomaly (which is due to the gluons and is responsible for OZI
violations) which is responsible for the scale dependence in $a^0(Q^2)$ 
and $\D g(Q^2)$ whereas (in the AB scheme) $\D\S$ is scale invariant.
The essence of this conjecture will reappear in the next section where
we describe the CPV method.

To test this conjecture, we need to find a way to measure $\D g(Q^2)$
itself, rather than the combination $a^0(Q^2)$. One possibility[16,18]
is to exploit the different scaling behaviours
of $\D q^S(x)$ and $\D g(x,Q^2)$ to distinguish their contributions in 
measurements of $g_1^p(x,Q^2)$ at different values of $Q^2$.
A second is to extract $\D g(x,Q^2)$ from processes such as open charm
production, $\c^* g \ra c \bar c$, which will be studied in various
forthcoming experiments at COMPASS, RHIC, etc.

\vskip1cm
\noindent{\bf 4.~~The CPV Method and Topological Charge Screening}
\vskip0.5cm
In this section, we shall discuss a less conventional approach 
to DIS based on a decomposition of matrix elements into products of
composite operator propagators and their associated 1PI vertex
functionals. This formalism has been developed in a series of papers[4-6,13]
on the `proton spin' problem. The starting point, as indicated above, 
is the use of the OPE in the proton matrix element of two currents.
This gives the standard form for a generic structure function moment:
$$
\int_0^1 dx~ x^{n-1} F(x;Q^2) = \sum_i C_i^n(Q^2) \langle p|\OO_i^n(0)
|p\rangle
\eqno(4.1)
$$
where $\OO_i^n$ are the set of lowest twist, spin $n$ operators in the OPE
and $C_i^n(Q^2)$ the corresponding Wilson coefficients. 
In the CPV approach, we now factorise the matrix element into the product 
of composite operator propagators and vertex functions, as illustrated in
Fig.~4.
\vskip0.2cm
\centerline{
{\epsfxsize=3.5cm\epsfbox{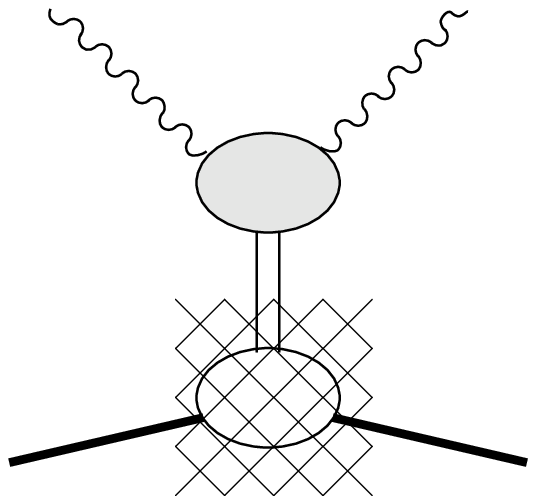}}}
\noindent{\eightrm Fig.4~~CPV description of DIS. The double line denotes
the composite operator propagator and the lower blob the 1PI vertex.}
\vskip0.2cm
To do this, we first select a set of composite operators $\tilde {\OO}_i$
appropriate to the physical situation and define vertices 
$\C_{\tilde{\OO}_i pp}$ as 1PI with respect to this set.
Formally, this is achieved by introducing sources for 
these operators in the QCD generating functional $W$, then performing a 
(partial) Legendre transform[5] to obtain a generating functional 
$\C[\tilde{\OO}_i]$. The 1PI vertices are the functional derivatives 
of $\C[\tilde{\OO}_i]$. The generic structure function sum rule (4.1) 
then takes the form
$$\eqalignno{
\int_0^1dx~x^{n-1}~F(x,Q^2) &=  
\sum_i \sum_j C_j^{(n)}(Q^2) \langle0|T~\OO_j^{(n)} ~
\tilde{\OO}_i |0\rangle \C_{\tilde{\OO}_i pp} \cr
&= \sum_i \sum_j C_j P_{ji} V_i
&(4.2) \cr }
$$
in a symbolic notation.

This decomposition splits the structure function into three pieces -- first,
the Wilson coefficients $C_j^{(n)}(Q^2)$ which control the $Q^2$ dependence 
and can be calculated in perturbative QCD; second, non-perturbative but
{\it target-independent} QCD correlation functions 
$\langle0|T~\OO_j^{(n)} ~\tilde{\OO}_i |0\rangle$; and third,
non-perturbative, target-dependent vertex functions $\C_{\tilde{\OO}_i pp}$
describing the coupling of the target proton to the composite operators 
of interest.
The vertex functions cannot be calculated directly from first principles.
They encode the information on the nature of the proton state and play an 
analogous role to the parton distributions in the more conventional
parton picture. 

It is important to recognise that this
decomposition of the matrix elements into products of propagators
and proper vertices is {\it exact}, independent of the choice of
the set of operators $\tilde{\OO}_i$. In particular, it is not necessary
for $\tilde{\OO}_i$ to be in any sense a complete set. All that happens if a 
different choice is made is that the vertices $\C_{\tilde{\OO}_i pp}$
themselves change, becoming 1PI with respect to a different
set of composite fields. Of course, while any set of $\tilde{\OO}_i$ may be
chosen, some will be more convenient than others. Clearly, the set 
of operators should be as small as possible while still capturing the
essential physics (i.e.~they should encompass the relevant degrees of
freedom) and indeed a good choice can result in vertices $\C_{\tilde{\OO}_i pp}$
which are both RG invariant and closely related to low energy physical 
couplings, such as $g_{\p NN}$. In this case, 
eq.(4.2) provides a rigorous relation between high $Q^2$ DIS and low-energy 
meson-nucleon scattering. 

For the first moment sum rule for $g_1^p$, it is most convenient 
to use the $U_A(1)$ anomaly equation immediately to re-express $a^0(Q^2)$ 
in terms of the forward matrix element of the topological charge $Q$, i.e.
$$
a^0(Q^2) ~=~{1\over 2M} 2n_f \langle p|Q|p\rangle
\eqno(4.3)
$$
where $M$ is the nucleon mass.

Our set of operators $\tilde{\OO}_i$ is then chosen to be the renormalised 
flavour singlet pseudoscalars $Q$ and $\Phi_5$, where 
$\Phi_5$ is simply the operator $\phi_5^0$ of section 2
with a special, and crucial, normalisation.
The normalisation factor is chosen such that in the OZI limit of QCD, where 
the anomaly is absent, $\Phi_{5}$ would have the correct normalisation to 
couple with unit decay constant to the $U_A(1)$ Goldstone boson which would 
exist in this limit. 
This also ensures that the vertex is RG scale independent. (The proof 
may be found in refs.[5,13].) The vertices are defined from the 
generating functional $\C[S_\m^a, Q, \phi_5^a, \phi^a]$ where
$$
\C[S_\m^a, Q, \phi_5^a, \phi^a] = W[S_\m^a, \o, S_5^a, S^a] - 
\int dx~\Bigl(\o Q + S_5^a \phi_5^a + S^a \phi^a \Bigr) 
\eqno(4.4)
$$
We then have
$$
\C_{1~singlet}^p ={1\over9} {1\over2M} 2n_f
C_1^{S}(\a_s)   
\biggl[\langle 0|T~Q~ Q|0\rangle \hat\C_{Qpp}
+\langle 0|T~ Q~ \Phi_{5}|0\rangle \hat\C_{\Phi_5 pp} \biggr]
\eqno(4.5)
$$
where the propagators are at zero momentum and the vertices 
are 1PI w.r.t.~$Q$ and $\Phi_{5}$ only. 
For simplicity, we have also introduced the notation
$i\bar u \C_{Qpp}u = \hat\C_{Qpp} \bar u \c_5 u$, etc.
This is illustrated in Fig.~5.
\vskip0.3cm
\centerline{
{\epsfxsize=8cm\epsfbox{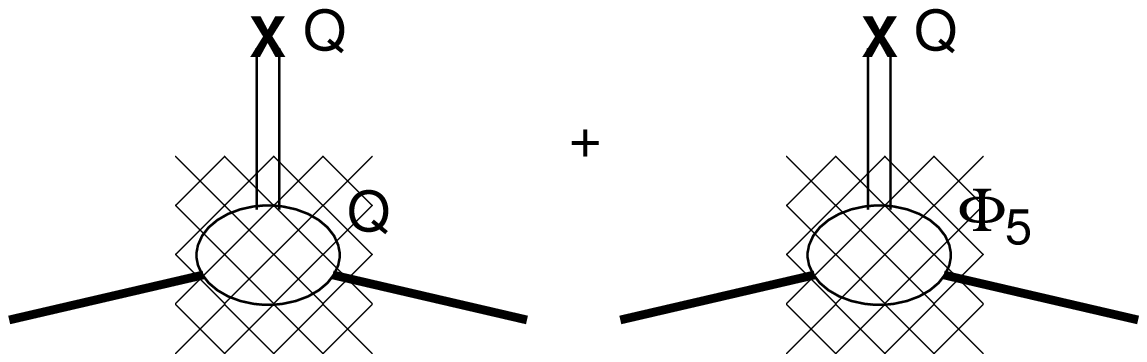}}}
\vskip0.3cm
\noindent{\eightrm Fig.5~~CPV decomposition of the matrix element 
$\langle p|Q|p\rangle$. The propagator in the first diagram is
$\chi(0)$; in the second it is $\sqrt{\chi'(0)}$.}
\vskip0.2cm
The composite operator propagator in the first term is simply 
the (zero-momentum) QCD topological susceptibility $\chi(0)$ which,
as we have seen in section 2, vanishes for QCD with massless quarks.
Furthermore, with the normalisation specified above for $\Phi_5$, the  
propagator $\langle 0|T~ Q~\Phi_5|0\rangle$ at zero momentum
is simply the square root of the slope of the topological
susceptibility. 

To see this, notice that by virtue of their definition in terms of
the generating functional (4.4), the matrix of 2-point vertices in the 
pseudoscalar sector is simply the inverse of the corresponding matrix
of pseudoscalar propagators, i.e.
$$
\left(\matrix{\C_{QQ} &\C_{Q\phi_5^0} \cr 
\C_{\phi_5^0 Q} &\C_{\phi_5^0 \phi_5^0} \cr }\right) ~=~ -
\left(\matrix{W_{\o\o} & W_{\o S_5^0} \cr
W_{S_5^0\o} & W_{S_5^0 S_5^0} \cr }\right)^{-1}
\eqno(4.6)
$$
This implies 
$$
\C_{\phi_5^0 \phi_5^0} = - W_{\o\o} \bigl({\rm det}~
W_{{\cal S}{\cal S}}\bigr)^{-1}
\eqno(4.7)
$$
letting ${\cal S}$ represent the set $\{\o, S_5^0\}$.
Differentiating w.r.t.~$k^2$ and taking the limit $k^2 = 0$,  
exploiting the fact that $W_{\o\o}(0)$ vanishes, we find
$$
{d\over dk^2}\C_{\phi_5^0 \phi_5^0}\big|_{k=0} =
\chi'(0) W_{\o S_5^0}^{-2}(0)
\eqno(4.8)
$$
Finally, normalising the field $\Phi_5$ proportional to $\phi_5^0$ such 
that ${d\over dk^2}\C_{\Phi_5 \Phi_5}\big|_{k=0} = 1$ (see refs.[5,13]
for further discussion), we find the 
required relation
$$
\langle0|T~Q~\Phi_5|0\rangle\big|_{k=0} = \sqrt{\chi'(0)}
\eqno(4.9)
$$
We therefore find:
$$
\C_{1~singlet}^p ~=~ {1\over9} {1\over 2M} 2n_f~
C_1^{S}(\a_s) ~\sqrt{\chi^{\prime}(0)} ~\hat\C_{\Phi_{5} pp}
\eqno(4.10)
$$

The slope of the topological susceptibility $\chi'(0)$ is
not RG invariant but, as shown in eq.(2.12), scales with the anomalous 
dimension $2\c$, i.e.
$$
{d\over dt} \sqrt{\chi'(0)} = \c \sqrt{\chi'(0)}
\eqno(4.11)
$$
On the other hand, the proper vertex has been chosen specifically
so as to be RG invariant[5,13]. The renormalisation group 
properties of this decomposition are crucial to our proposed resolution 
of the `proton spin' problem.
 
Our proposal[4,5] is that we should expect the source of OZI 
violations to lie in the RG non-invariant, and therefore anomaly-sensitive, 
terms, i.e.~in $\chi^{\prime}(0)$ rather than in the RG invariant vertex.
This is the key assumption that allows us to make a quantitative prediction
for $\C_1^p$ on the basis of a calculation of the topological susceptibility
alone. The RG invariance of the vertices is a necessary condition for this
assumption to be reasonable. Further phenomenological evidence from $U_A(1)$
current algebra supporting this conjecture is discussed in refs.[5,19].
Notice that we are using RG non-invariance, i.e.~dependence 
on the anomalous dimension $\c$, merely as an indicator of which quantities
are sensitive to the anomaly and therefore likely to show OZI violations.
Since the anomalous suppression in $\C_1^p$ is thus assigned to the composite 
operator propagator rather than the proper vertex, the suppression is a 
{\it target independent} property of QCD related to the anomaly, 
not a special property of the proton structure.

In this picture, therefore, the basic dynamical mechanism 
responsible for the suppression of the `proton spin' can be identified
as {\it topological charge screening} by the QCD vacuum. That is,
when a matrix element of the topological charge is
measured, the QCD vacuum screens the topological charge through
the zero or anomalously small values of the susceptibility
$\chi(0)$ and its slope $\chi'(0)$ respectively (see Fig.~4).
The mechanism is analogous to the screening of electric charge in QED.
There, because of the gauge Ward identity, the screening is given entirely
by the (`target independent') dressing of the photon propagator by
vacuum polarisation diagrams, leading to the relation $e_R = e_B
\sqrt{Z_3}$ (with $Z_3<1$) between the renormalised and bare charges,
in direct analogy to eq.(4.12) below with the topological susceptibility
playing the role of the photon propagator.

To convert this into a quantitative prediction we use the OZI 
approximation\footnote{${}^{(5)}$}{\eightpoint
\noindent Without this approximation, our result (4.12) reads:
$$
{a^0(Q^2)\over a^8} = \biggl({\sqrt6\over f_{\pi}} 
\sqrt{\chi'(0)}|_{Q^2}\biggr)~\biggl( 
{\hat\C_{\Phi_5 pp} \over \sqrt2 \hat\C_{\Phi_5^8 pp}}\biggr) ~\equiv~
s_P s_V
$$
A (target-independent) propagator suppression factor $s_P < 1$
is a signal of topological charge screening; a (target-dependent)
vertex suppression factor $s_V < 1$ would be a sign that the  
proton coupling to the flavour singlet Goldstone boson $\eta_{OZI}^o$
in the OZI limit of QCD is small, equivalent in parton terms to
a significant (negatively polarised) strange quark content in the proton.
Of course both effects may be present. However, our conjecture is that
the `proton spin' problem is best explained by $s_P < 1$, 
$s_V \simeq 1$.} for the vertex  $\hat\C_{\Phi_{5} pp}$. In terms of a 
similarly normalised octet field $\Phi_5^8$, this is
$\hat\C_{\Phi_5 pp} = \sqrt2 \hat\C_{\Phi_5^8 pp}$. 
The normalisation is crucial in allowing the use of the
OZI relation here. The corresponding OZI 
prediction for $\sqrt{\chi'(0)}$ would be $f_{\pi}/\sqrt6$. These OZI 
values are determined by comparing the result (4.10) (at least that part
relating to the proton matrix element) with the conventional
Goldberger-Treiman relation for the flavour octet axial charge in 
the chiral limit (see refs.[5,13]). This gives our key formula\footnote
{${}^{(6)}$}{\eightpoint
\noindent It is interesting to compare eq.(4.12) with the well-known
Witten-Veneziano formula[20,21] which, at leading order in the $1/N_c$
expansion, relates the mass of the $\eta'$ (in massless QCD) to 
$\chi^{YM}$, the topological susceptibility of pure gluodynamics:
$$
m_{\eta'} \simeq {\sqrt6\over f_\pi} \sqrt{\chi^{YM}(0)}
$$}
$$
{a^0(Q^2)\over a^8} = {\sqrt6\over f_{\pi}} \sqrt{\chi'(0)}\big|_{Q^2}
\eqno(4.12)
$$
for the flavour singlet axial charge. Incorporating this into
the formula for the first moment of the polarised structure function, 
we find
$$
\C_{1~singlet}^p ~=~ {1\over9} C_1^{S}(\a_s) ~a^8~
{\sqrt6\over f_{\pi}}~\sqrt{\chi'(0)}
\eqno(4.13)
$$

The final step involves an explicit calculation of $\chi'(0)$. 
This was done in ref.[6] using the QCD spectral sum rule (QSSR) method.
We find[6,13]
$$
\sqrt{\chi'(0)} = (26.4 \pm 4.1) ~\MV
\eqno(4.14)
$$
Substituting this in eq.(4.13), and also using the QSSR 
prediction for $f_\pi$ in order to minimise systematic errors, 
we arrive at our prediction for the first moment of the polarised structure
function in the chiral limit:
$$\eqalignno{
a^0(Q^2=10\GV^2) &= 0.33 \pm 0.05 \cr
\C_1^p(Q^2=10\GV^2) &= 0.144 \pm 0.009
&(4.15) \cr}
$$
This can be immediately compared with the Ellis-Jaffe sum rule prediction
$$
a^0\big|_{\rm EJ} = 0.58 \pm 0.03
\eqno(4.16)
$$
We therefore find a suppression of $a^0(Q^2)$ relative to the OZI
expectation by a factor of around 0.55.

\vskip0.3cm
A complete description of our calculation of the slope of the topological
susceptibility can be found in refs.[6,13].
The QSSR analysis of $\chi'(0)$ in the chiral limit of QCD
is essentially straightforward and shows a clear range of stability 
with respect to the Laplace sum rule parameter $\t$, indicating that
the prediction is reliable. The stabilisation scale,
$\t^{-1} \sim 2.5 - 5~\GV^2$, is sufficiently big for
higher dimensional condensates to be suppressed, and the calculation
displays the hierarchy of gluonium to light meson hadronic scales
anticipated by ref.[22].\footnote{${}^{(7)}$}{\eightpoint
\noindent Nevertheless, our application of QSSR to the $U_A(1)$ sector 
of QCD has been 
criticised repeatedly by Ioffe (see e.g.~refs.[23,24]) on two grounds:~ 
(i)~that there are important neglected contributions from `instantons', 
i.e.~higher dimensional
condensates, and ~(ii)~ that when the strange quark mass is included, QSSR
results for current-current correlators show quite unrealistic $SU(3)$ breaking.
Neither criticism is valid, and both are refuted in detail in our recent
paper[13]. We have commented in the text how the size of the stabilisation 
scale is sufficient to suppress higher dimensional condensates. 
Ioffe's second criticism is based on a 
calculation[25] where radiative corrections are not properly implemented
and a number of other errors are made. The correct results are given in ref.[13],
where we extend our previous analysis of the `proton spin' problem
systematically beyond the chiral limit
using a new set of generalised Goldberger-Treiman relations.}

Lattice gauge theory methods may also be used to calculate the topological
susceptibility. However, $\chi'(0)$ is a particularly difficult correlation 
function to calculate on the lattice, requiring algorithms that implement 
topologically non-trivial configurations in a sufficiently fast and 
efficient way.
Very preliminary results from the Pisa group[26] of calculations
in full QCD with dynamical quarks indicate a value of the order
$\sqrt{\chi'(0)} \sim (19 \pm 4) ~{\rm MeV}$. Given the preliminary nature
of the lattice simulations, the rough agreement with the QSSR result is
encouraging. 
Qualitative explanations of topological charge screening and the
anomalously small value of $\chi'(0)$ in QCD may also be given using models
of the instanton vacuum, e.g.~the instanton liquid model of ref.[27].

\vskip0.3cm
Finally, to complete this discussion, it is useful to recognise the 
complementary nature of the QCD parton model and the CPV method presented 
here. Both involve at present incalculable non-perturbative functions
describing the proton state -- the quark and gluon distributions in the 
parton picture and the 1PI vertices in the CPV method. Both exhibit a degree
of universality -- the same parton distributions may be used in different
QCD processes such as DIS or hadron-hadron collisions, while the vertices
(when they can be identified with low-energy couplings) also provide a link
between high $Q^2$ DIS and soft meson-nucleon interactions.

The principal attribute of the parton model is that it allows a detailed
description of the structure of the proton in terms of its quark and gluon
constituents. On the other hand,
one of the main advantages of the CPV method is that some non-perturbative 
information which is generic to QCD, i.e.~independent of the target, is
factored off into the composite operator propagator. This allows us to 
distinguish between non-perturbative mechanisms which are generic to all
QCD processes and those which are specific to a particular target.
As explained above, our contention is that the anomalous suppression in the 
first moment of $g_1^p$ is of the first, target-independent, type.
This conjecture could in principle be tested by
DIS with non-nucleon targets, which may effectively be realised in
semi-inclusive polarised DIS. This will be discussed in section 6.

Both the parton and CPV methods allow a natural conjecture in which the 
origin of this suppression is attributed to `glue' -- either through a 
large polarised gluon distribution $\D g(Q^2)$ in the parton description or 
due to an anomalous suppression of the slope of the topological 
susceptibility $\chi'(0)$ in the CPV description. These conjectures 
are based on assumptions that the appropriate RG invariant quantities, 
$\D q^S$ or $\hat\C_{\Phi_5 pp}$, obey the OZI rule.
The motivation for this is particularly strong in the CPV case, where it is
supported by a range of evidence from low-energy meson phenomenology
in the $U_A(1)$ channel. Moreover, it identifies a fundamental physical
mechanism as being at the origin of the `proton spin' suppression -- the
screening of topological charge by the QCD vacuum.

The two approaches therefore provide related, but complementary, insights 
into the nature of the `proton spin' effect. Clearly, both insights are 
needed and both methods have a full part to play in understanding this 
intriguing and subtle phenomenon.

\vskip1cm

\noindent{\bf 5.~~Experiment}
\vskip0.5cm
In the last year, the SMC collaboration have completed their analysis
of the final data from the 1996 run. This is shown in Fig.~6.
\vskip0.2cm
\centerline{
{\epsfxsize=5.5cm\epsfbox{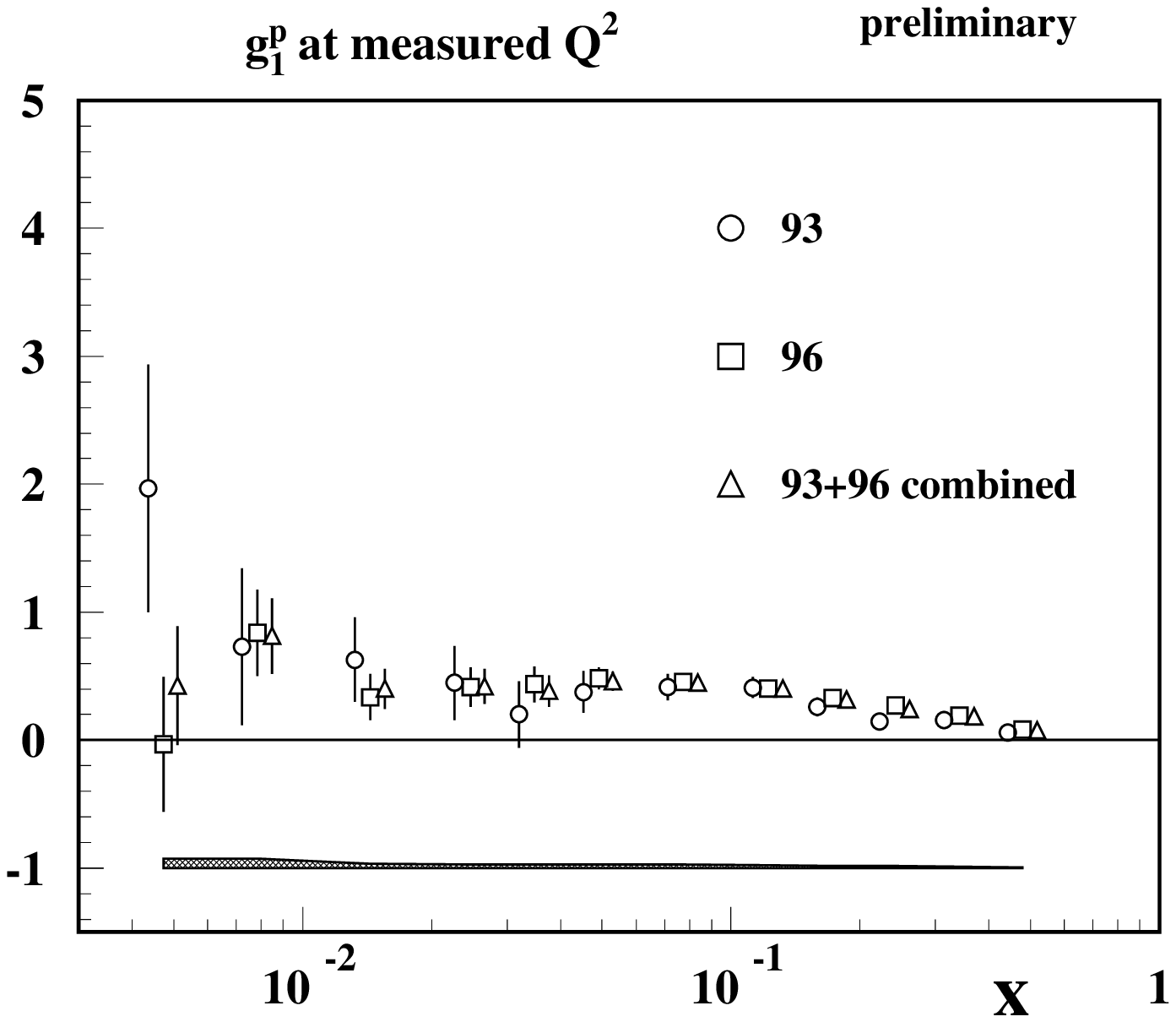}}}
\noindent{\eightrm Fig.6~~SMC data for $g_1^p$}
\vskip0.2cm
Including an uncontroversial extrapolation in the unmeasured region
$0.7<x<1$, and taking into account the RG scaling from the measured
$Q^2$ for a particular $x$ to the reference value $Q^2 = 10~\GV^2$, SMC 
quote[28] the following result for the first moment of $g_1^p$ in the
measured range $x>0.003$:
$$
\C_1^p (Q^2=10\GV^2)\big|_{x>0.003} \equiv \int_{0.003}^1 dx~g_1^p(x;Q^2)
= 0.141 \pm 0.006 \pm 0.008 \pm 0.006
\eqno(5.1)
$$
where the first error is statistical, the second is systematic and the
third is due to the uncertainty in the $Q^2$ evolution.
\vskip0.2cm
\centerline{
{\epsfxsize=5cm\epsfbox{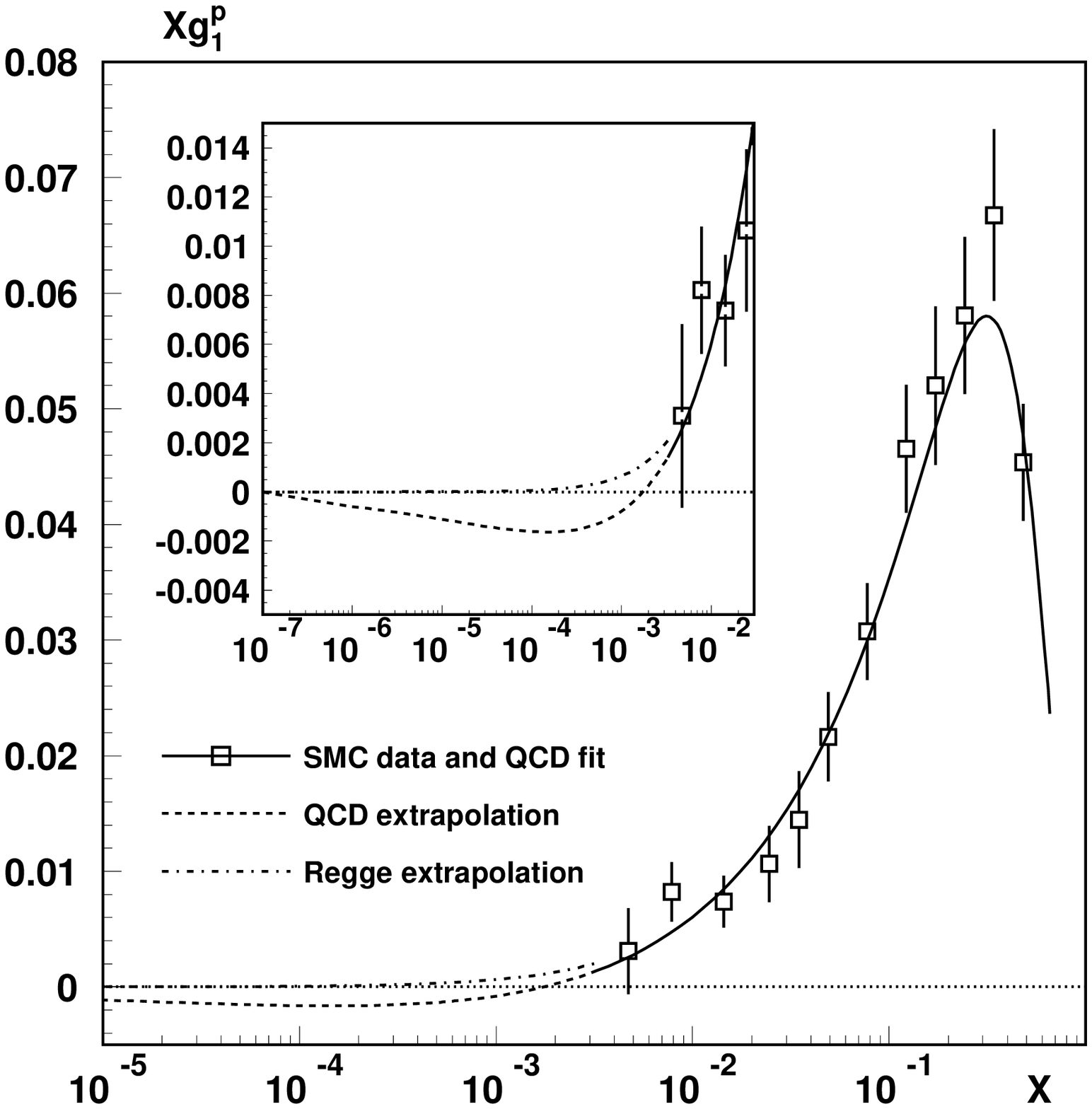}}}
\vskip0.4cm
\noindent{\eightrm Fig.7~~SMC data for $xg_1^p$ including
`Regge' and `QCD' small $x$ extrapolations.}
\vskip0.2cm

The result for the entire first moment depends on how the extrapolation to
the unmeasured small $x$ region $x<0.003$ is performed. See Fig.~7.
This is still a controversial issue. Using a simple
Regge fit, SMC find $\C_1^p = 0.142 \pm 0.017$ from which they deduce
$a^0= 0.34 \pm 0.17$. Alternatively, using a small $x$ fit 
in which the parton densities are parametrised at small $x$ at a low
$Q^2$ scale and then extrapolated to higher $Q^2$ using the perturbative
QCD evolution equations[16,29], SMC quote
$\C_1^p = 0.130 \pm 0.017$ and
$a^0 = 0.22 \pm 0.17$ ~(all at $Q^2 = 10\GV^2$). 
The key feature of the evolution equations is that while
$\D\S(x,Q^2)$ falls with increasing $Q^2$, the polarised gluon distribution
$\D g(x,Q^2)$ rises. The net effect is that $g_1^p(x,Q^2)$ is driven strongly
negative at $Q^2=10\GV^2$ for sufficiently small $x$.
This gives a potentially large negative contribution to the first moment
$\C_1^p$, but with relatively large errors[29].

Moreover, SMC have recently published an alternative analysis[30] of their 
data, using a different event selection, this time quoting a slightly lower 
number for the integral over the measured region of $x$:
$$
\C_1^p (Q^2=10\GV^2)\big|_{x>0.003} \equiv \int_{0.003}^1 dx~g_1^p(x;Q^2)
= 0.133 \pm 0.005 \pm 0.006 \pm 0.004
\eqno(5.2)
$$

Clearly it is premature to draw too strong a conclusion given the
large errors on the experimental determinations of $\C_1^p$ and
$a^0$ and the uncertainty over the small $x$ extrapolation.
Future studies of the small $x$ region of $g_1^p$ and the
$Q^2$ scaling behaviour of the gluon distribution $\D g(x,Q^2)$
will be important challenges to experimentalists.
Nevertheless, it is extremely encouraging that the CPV prediction (4.15)
is firmly in the region favoured by the data. This gives us extra
confidence that the explanation of the `proton spin' problem in terms of 
topological charge screening by the QCD vacuum is correct.

\vfill\eject
%\vskip1cm

\noindent{\bf 6.~~Semi-Inclusive Polarised DIS}
\vskip0.5cm

Recently, a new proposal to exploit semi-inclusive DIS in the target
fragmentation region to elucidate the `proton spin' effect has been 
presented[10,11]. The idea is to test the mechanism of topological 
charge screening, or more precisely the prediction of `target independence',
suggested by the CPV method by using semi-inclusive DIS in effect to make
measurements of the polarised structure functions of other hadronic targets
besides the proton and neutron.

The essence of the target independence conjecture
is that for any hadron, the singlet
axial charge in eq.(2.3) can be substituted by its OZI value multiplied by a 
universal (target-independent) suppression factor $s(Q^2)$ determined, up to 
radiative corrections, by the anomalous suppression of the first moment
of the topological susceptibility $\sqrt{\chi'(0)}$. For example,
for a hadron containing only $u$ and $d$ quarks, the OZI relation
is simply $a^0 = a^8$, so we predict:
$$
\C_1 = {1\over12} C_1^{NS} \Bigl( a^3 + {1\over3}(1 + 4s) a^8 \Bigr)
\eqno(6.1)
$$
where
$$
s(Q^2) = {C_1^{S}(\a_s) \over C_1^{NS}(\a_s)}~{a^0(Q^2)\over a^8}
\eqno(6.2)
$$
Since $s$ is target independent, we can use the value measured for the proton 
to deduce $\C_1$ for any other hadron simply from the flavour 
non-singlet axial charges.
From our spectral sum rule estimate of $\sqrt{\chi'(0)}$, we find 
$s\sim 0.6$ at $Q^2=10~{\rm GeV}^2$, while the central values of 
the results for $a^0(Q^2)$ extracted by SMC from the data give 
even lower values for $s$. 

The non-singlet axial charges for a hadron $\BB$ are given by the matrix
elements of the flavour octet axial currents, so can be factorised
into products of $SU(3)$ Clebsch-Gordon coefficients and reduced matrix
elements. Together with the target independence conjecture, this allows
predictions to be made for ratios of the first moments of the
polarised structure functions $g_1^{\BB}$ for different $\BB$, which involve
only group theoretic numbers and the universal suppression factor $s$.
Some of the most interesting are:
$$
\C_1^p / \C_1^n ~~=~~ {2s -1 -3(2s+1)F/D \over 2s +2 -6sF/D} 
\eqno(6.3)
$$
$$
\C_1^{\D^{++}} / \C_1^{\D^-} ~~=~~ \C_1^{\S_c^{++}} / \C_1^{\S_c^0} ~~=~~
{2s+2 \over 2s-1} 
\eqno(6.4)
$$
where $\S_c^{++}$ ~($\S_c^0$) is the state with valence quarks $uuc$~
($ddc$).
The results for $\D$ and $\S_c$ are particularly striking because of the 
$2s-1$ denominator factor, which is very small for the range of $s$ favoured by
experiment. These examples therefore show spectacular deviations from the 
valence quark counting (OZI) expectations, which would give the ratio 4.
They also turn out to be the ones which allow the most clear-cut
experimental interpretation.

The proposal of ref.[10] is that these ratios can be realised in 
semi-inclusive DIS in a kinematical region where the detected hadron
$h$ (a pion or {$D$} meson in these examples) carries a large target
energy fraction, i.e.~$z$ approaching 1, with a small invariant 
momentum transfer. To understand this, we briefly
review some of the theory of semi-inclusive DIS.

\vskip0.2cm
\centerline{
{\epsfxsize=2.8cm\epsfbox{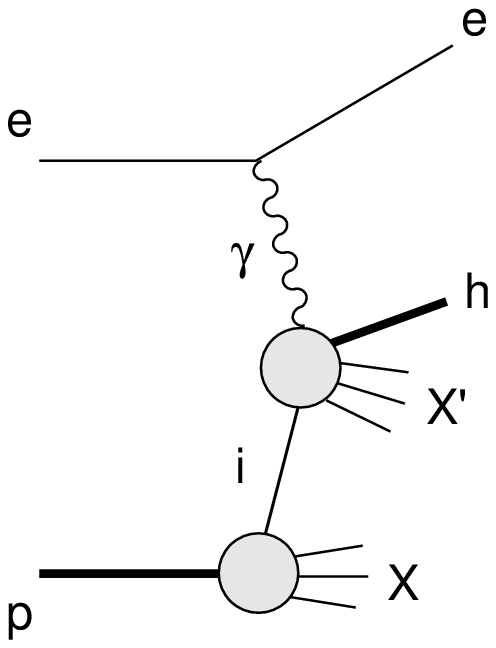}}}
\noindent{\eightrm Fig.8~~Semi-inclusive DIS: current fragmentation region.}
\vskip0.0cm
\centerline{
{\epsfxsize=2.8cm\epsfbox{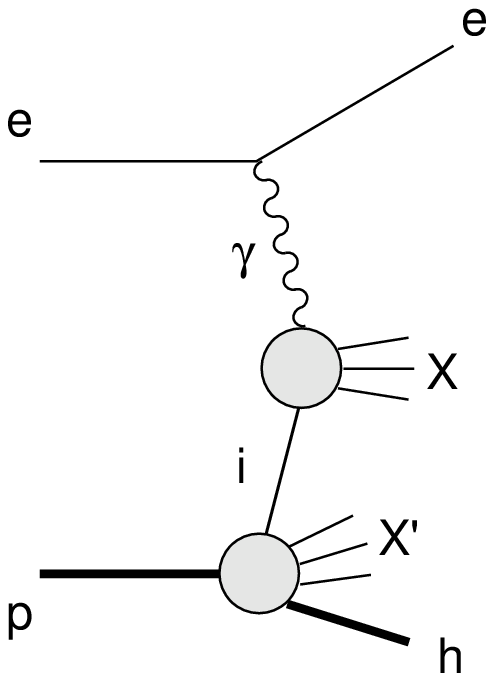}}}
\noindent{\eightrm Fig.9~~Semi-inclusive DIS: target fragmentation region.}
\vskip0.2cm

There are two distinct contributions to the semi-inclusive DIS reaction 
$e N \ra e h X$, coming from the current and target fragmentation regions. 
These are shown in Figs.~8 and 9.
The current fragmentation events are described by parton fragmentation
functions $D_i^h({\tilde z\over 1-x},Q^2)$, where $i$ denotes the parton, while 
the target fragmentation events are described by fracture functions[31]
$M_i^{hN}(x,\tilde z,Q^2)$ representing the joint probability distribution for
producing a parton with momentum fraction $x$ and a detected hadron $h$ 
(with momentum $p_2^{\prime}$) carrying energy fraction $\tilde{z}$ from a 
nucleon $N$.
The lowest order cross section for polarised semi-inclusive DIS is:
$$
x~{d\D\s \over dx dy d\tilde z} = {Y_P\over2} {4\pi\a^2 \over s} \sum_i 
e_i^2 \biggl[\D M_i^{hN}(x,\tilde z,Q^2) 
+ {1\over 1-x} \D q_i(x,Q^2) D_i^h\Bigl({\tilde z\over 1-x}, Q^2\Bigr)
\biggr]
\eqno(6.5)
$$
where $\tilde z = E_h/E_N$ (in the photon-nucleon CM frame).
The notation is slightly different from section 3. Here, $\D q_i(x)$
refers to quarks and antiquarks separately and a sum over both is implied.
$\D M_i^{hN}(x,\tilde z,Q^2)$ is the polarisation asymmetry of the 
fracture function. The NLO corrections to eq.(6.5) are given in ref.[32].

In fact, for our purposes it is better to use the extended fracture 
functions introduced recently in refs.[33]. These new fracture functions 
$\D M_i^{hN}(x,z,t, Q^2)$ have an explicit dependence on the 
invariant momentum transfer $t$. The original fracture functions are
found by integrating over $t$ in the range $t<O(Q^2)$. One of the
advantages of the extended fracture functions is that they obey
a simple, homogeneous, RG evolution equation:
$$
{\pl\over\pl\ln Q^2} \D M_i^{hN}(x,z,t, Q^2) =
{\a_s\over2\pi} \int_x^{1-z} {d\w\over \w} 
\D P_{ij}\Bigl({x\over\w},\a_s\Bigr) \D M_j^{hN}(\w,x,t, Q^2)
\eqno(6.6)
$$
The differential cross section in the target fragmentation can then
be written, analogously to eq.(2.1), as
$$
x{d \D \s^{target}\over dx dy dz dt} = {Y_P\over2}{4\pi\a^2\over s}
\D M_1^{hN}(x,z, t, Q^2)
\eqno(6.7)
$$
where $\D M_1^{hN}$ is the fracture function equivalent of the
inclusive structure function $g_1^N$, (c.f.~eq.(3.2)):
$$
\D M_1^{hN}(x,z, t, Q^2) = {1\over2}\sum_i e_i^2
\D M_i^{hN}(x,z, t, Q^2) 
\eqno(6.8)
$$

For $z$ approaching 1, i.e.~the hadron carrying a large 
target energy fraction, this target fragmentation process may be 
simply modelled by a single Reggeon exchange (see Fig.~10). This
corresponds to the approximation
$$
\D M_1^{hN}(x,z, t, Q^2) ~\SIMEQz~ F(t) (1-z)^{-2\a_{\BB}(t)}
g_1^{\BB}(x_{\BB},t,Q^2)
\eqno(6.9)
$$
The notation here is $z=p_2^{\prime}.q/p_2.q$, $x_{\BB} = Q^2/2k.q$,
$1-z = x/x_{\BB}$, and $t = -k^2 \ll O(Q^2)$ so that $z\simeq\tilde z$.
The Reggeon emission factor $F(t)(1-z)^{-2\a_{\BB}(t)} $ cancels in 
the ratios of cross sections we consider.
\vskip0.2cm
\centerline{
{\epsfxsize=2.8cm\epsfbox{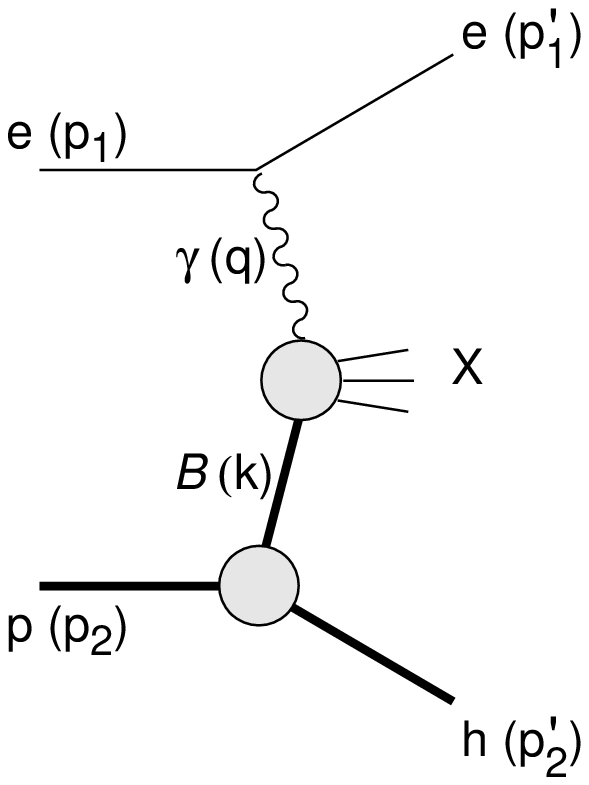}}}
\noindent{\eightrm Fig.10~~Single Reggeon exchange model of 
$ep\ra e h X$.}
\vskip0.2cm

Although single Reggeon exchange is only an approximation to the more
fundamental QCD description in terms of fracture functions, it shows
particularly clearly how observing semi-inclusive processes at large
$z$ with particular choices of $h$ and $N$ amounts in effect to
performing inclusive DIS on virtual hadronic targets $\BB$.
Since the predictions (6.3,6.4) depend only on the $SU(3)$ properties
of $\BB$, together with target independence, they will hold equally 
well when $\BB$ is interpreted as a Reggeon rather than a pure hadron 
state. For example, from the quark diagram in Fig.~11, we easily see
that the semi-inclusive reaction $ep\ra e\pi^- X$ measures $g_1^{\BB}$
for $\BB = \D^{++}$.
\vskip0.2cm
\centerline{
{\epsfxsize=4.3cm\epsfbox{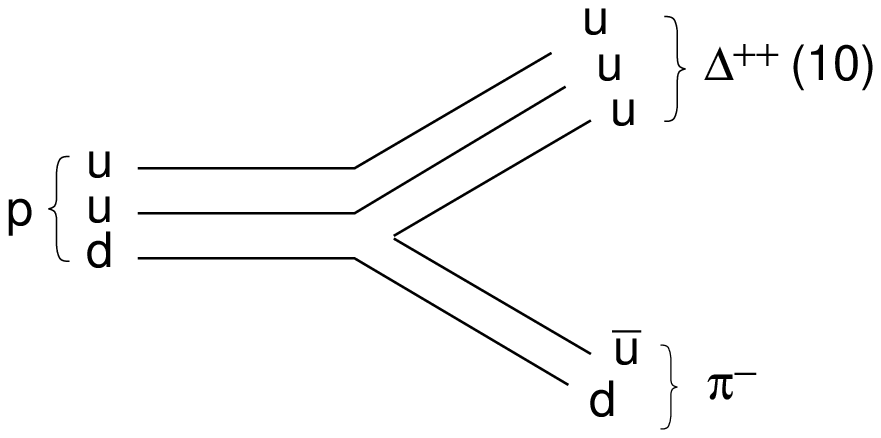}}}
\noindent{\eightrm Fig.11~~Quark diagram for the ${Nh}{\cal B}$ vertex in
the reaction $ep\ra e\pi^- X$ where ${\cal B}$ has the quantum numbers of
$\D^{++}$.}
\vskip0.2cm

We can therefore make simple predictions for the ratios ${\cal R}$ of the
first moments of the polarised fracture functions
$\int_0^{1-z} dx \D M_1^{hN}(x,z, t, Q^2)
\simeq \int_0^1 dx_{\BB} g_1^{\BB}(x_{\BB},t, Q^2)$ for various reactions.
We emphasise again that these do not depend on any detailed model
of the fracture functions, such as single or even multi-Reggeon exchange.
(A more formal justification in terms of the fracture function --
cut vertex equivalence[33] is currently under study.)
The ratios (6.3,6.4) are obtained in the limit as $z$ approaches 1, 
where the reaction $eN\ra e h X$ is dominated by the process
in which most of the target energy is carried through into the 
final state $h$ by a single quark. We therefore predict
$$
{\cal R} \biggl({ep\ra e\pi^- X\over en\ra e\pi^+ X}\biggr) ~\SIMEQz~
{2s+2\over2s-1}
\eqno(6.10)
$$
Similarly for charmed mesons:
$$
{\cal R} \biggl({ep\ra eD^- X\over en\ra eD^0 X}\biggr) ~\SIMEQz~
{2s+2\over2s-1}
\eqno(6.11)
$$
For strange mesons, on the other hand, the ratio depends on whether
the exchanged object has $SU(3)$ quantum numbers in the $\bf 8$ or 
$\bf 10$ representation, so the prediction is less conclusive, e.g.
$$\eqalignno{
{\cal R}\biggl({ep\ra eK^0 X\over en\ra eK^+ X}\biggr) ~\SIMEQz~~
&{2s-1-3(2s+1)F^*/D^* \over 2s-1-3(2s-1)F^*/D^*}~~~({\bf 8}) \cr
&{2s+1\over 2s-1}~~~({\bf 10}) 
&(6.12) \cr }
$$

At the opposite extreme, for
$z$ approaching 0, the detected hadron carries only a small fraction of the 
target nucleon energy. In this limit, the ratio of 
cross section moments for 
$ep\ra \el \pi^-(D^-)X$ and $en \ra e \pi^+(D^0)X$ 
is simply the ratio of the structure function moments for the proton and
neutron. 

Interpolating between these limits, we expect the ratio ${\cal R}$
in the range $0<z<1$ for the reactions $en \ra e \pi^+(D^0)X$ 
over $ep\ra \el \pi^-(D^-)X$ to look like the
sketch in Fig.~12.
\vskip0.2cm 
\centerline{
{\epsfxsize=5.5cm\epsfbox{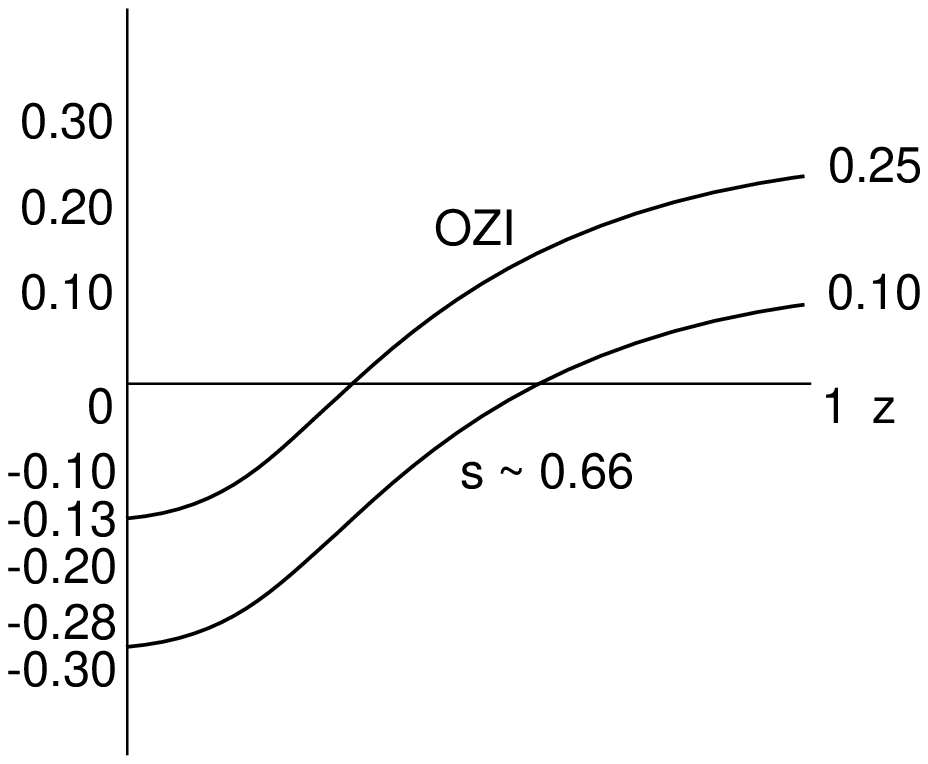}}}
\noindent{\eightrm Fig.12~~Cross section ratios ${\cal R}$ for 
$en\ra e\pi^+(D^0)X$ over $ep\ra e\pi^-(D^-)X$ between $z\ra 0$ 
and $z\ra 1$, contrasting the OZI and CPV predictions}
\vskip0.2cm
The difference between the OZI (or valence quark model) expectations 
and these predictions based on our target-independent interpretation
of the `proton spin' data is therefore quite dramatic, and should
give a clear experimental signal.

Since the proposed experiments require particle identification in the
target fragmentation region, they are difficult to do at a polarised 
fixed-target experiments such as COMPASS, or even HERMES, which are 
better suited to studying semi-inclusive processes in the current 
fragmentation region. A better option would be a polarised 
$ep$ collider, such as HERA[34]. To conclude, we shall make some brief
comments about the experimental requirements necessary for these
predictions to be tested[11].

The first requirement is to measure particles at extremely small
angles ($\theta \le 1$ mrad), corresponding to $t$ less than around
1 $\GV^2$. This has already been achieved at HERA in measurements of 
diffractive and leading proton/neutron scattering. The technique for
measuring charged particles involves placing detectors commonly known 
(but see the discussion session) as `Roman Pots' inside the beam pipe 
itself. In the present experiments, this detection system is known as the
Leading Proton Spectrometer (LPS).

The next point is to notice that the considerations above apply equally
to $\r$ as to $\pi$ production, since the ratios ${\cal R}$ are determined
by flavour quantum numbers alone. The particle identification (ID) 
requirements will therefore be less stringent, especially as the
production of leading strange mesons from protons or neutrons is strongly
suppressed. However, we require the forward detectors to have good 
acceptance for both positive and negatively charged mesons $M = \pi,\r$
in order to measure the ratio (6.10).

The reactions with a neutron target can be measured if the polarised
proton beam is replaced by polarised ${}^3 He$. In this case, if we assume
that ${}^3He = Ap + Bn$, the cross section for the production of positive
hadrons $h^+$ measured in the LPS is given by
$$
\s\bigl({}^3He \ra h^+\bigr) \simeq A \s\bigl(p\ra h^+\bigr)
+ B \s\bigl(n \ra p\bigr) + B\s\bigl(n\ra M^+\bigr)
\eqno(6.13)
$$
The first contribution can be obtained from measurements with the proton
beam. However, to subtract the second one, the detectors must have sufficient
particle ID at least to distinguish protons from positively charged mesons.

Finally, estimates of the total rates suggest that around $1\%$ of the 
total DIS events will contain a leading meson in the target fragmentation
region where the LPS has non-vanishing acceptance ($z>0.6$) and in
the dominant domain $x<0.1$. The relevant cross sections are therefore
sufficient to allow the ratios ${\cal R}$ to be measured.

We conclude that while these proposals undoubtedly
pose a challenge to experimentalists, they are nevertheless possible.
Given the theoretical importance of the `proton spin' problem
and the topological charge screening mechanism, there is therefore 
strong motivation to perform target fragmentation experiments at
polarised HERA.

\vskip0.6cm

\noindent{\bf Acknowledgements}
\vskip0.3cm
I would like to thank the directors of the school, Profs.~A.~Zichichi,
G.~'t Hooft and G.~Veneziano for the invitation to lecture in such
a wonderful location and the students for their interest and
enthusiasm. I am also grateful to D.~De Florian, S.~Narison and
G.~Veneziano for their collaboration on the original work presented
here. This was supported in part by the EC TMR Network Grant
FMRX-CT96-0008 and by PPARC.

\vfill\eject

\noindent{\bf References}
\vskip0.3cm

\settabs\+\ [&1] &Author \cr

\+\ [&1] &EMC Collaboration, J.~Ashman {\it et al.}, {\it Phys.~Lett.} B206 
(1988) 364; \cr
\+\ &{} &~~~~~~~~~{\it Nucl.~Phys.} B328 (1989) 1. \cr

\+\ [&2] &J.~Ellis and R.L.~Jaffe, {\it Phys.~Rev.} D9 (1974) 1444. \cr

\+\ [&3] &G.~Altarelli and G.G.~Ross, {\it Phys.~Lett.} B212 (1988) 391. \cr

\+\ [&4] &G.M.~Shore and G.~Veneziano, {\it Phys.~Lett.} B244 (1990) 75. \cr

\+\ [&5] &G.M.~Shore and G.~Veneziano, {\it Nucl.~Phys.} B381 (1992) 23. \cr

\+\ [&6] &S.~Narison, G.M.~Shore and G.~Veneziano, {\it Nucl.~Phys.}
B433 (1995) 209.\cr

\+\ [&7] &G.~Veneziano, {\it Mod.~Phys.~Lett.} A4 (1989) 1605. \cr

\+\ [&8] &S.L.~Adler, {\it Phys.~Rev.} 177 (1969) 2426; \cr
\+\ &{}  &J.S.~Bell and R.~Jackiw, {\it Nuovo Cim.} 60 (1969) 47.\cr

\+\ [&9] &G.M.~Shore, {\it in} Proceedings of the 1998 Zuoz Summer School\cr
\+\ &{} &~~~~~~~~~~on {\it Hidden Symmetries and Higgs Phenomena},
hep-ph/9812xxx. \cr

\settabs\+ [1&1] &Authors \cr

\+ [1&0] &G.M.~Shore and G.~Veneziano, {\it Nucl.~Phys.} B516 (1998) 333. \cr

\+ [1&1] &D.~De Florian, G.M.~Shore and G.~Veneziano, {\it in} Proceedings 
of 1997 Workshop  \cr
\+\ &{} &~~~~~~~~~{\it Physics with Polarised Protons at} HERA, 
hep-ph/9711358. \cr

\+ [1&2] &D.~Espriu and R.~Tarrach, {\it Z.~Phys} C16 (1982) 77. \cr

\+ [1&3] &S.~Narison, G.M.~Shore and G.~Veneziano, hep-ph/9812333. \cr

\+ [1&4] &G.~Veneziano, {\it in `From Symmetries to Strings: Forty Years of
Rochester Conferences'}, \cr
\+\ &{} &~~~~~~~~~~ed. A.~Das, World Scientific, 1990. \cr 

\+ [1&5] &X.~Ji, {\it Phys.~Rev.~Lett.} 78 (1997) 610. \cr

\+ [1&6] &R.D.~Ball, S.~Forte and G.~Ridolfi, {\it Phys.~Lett.} B378 (1996)
255. \cr

\+ [1&7] &R.D.~Ball, {\it in} Proceedings, Ettore Majorana International 
School of Nucleon Structure,  \cr
\+\ &{} &~~~~~~~~~~Erice, 1995; hep-ph/9511330. \cr

\+ [1&8] &R.~Ball {\it et al.}, hep-ph/9609515. \cr

\+ [1&9] &G.M.~Shore and G.~Veneziano, {\it Nucl.~Phys.} B381 (1992) 3. \cr

\+ [2&0] &E.~Witten, {\it Nucl.~Phys.} B156 (1979) 269. \cr

\+ [2&1] &G.~Veneziano, {\it Nucl.~Phys.} B159 (1979) 213. \cr

\+ [2&2] &V.A.~Novikov {\it et al}, {\it Nucl.~Phys.} B237 (1984) 525.  \cr

\+ [2&3] &B.L.~Ioffe, {\it in} Proceedings, Ettore Majorana International
School of Nucleon Structure, \cr
\+\ &{} &~~~~~~~~~~Erice, 1995; hep-ph/9511401. \cr 

\+ [2&4] &B.L. Ioffe, Lecture at St.~Petersburg Winter School, hep-ph/9804328.\cr 

\+ [2&5] &B.L.~Ioffe and A.Yu.~Khodzhamiryan, {\it Yad.~Fiz.} 55 (1992) 3045. \cr

\+ [2&6] &G.~Boyd, B.~All\'es, M.~D'Elia and A.~Di Giacomo, {\it in} Proceedings,
HEP 97 Jerusalem, \cr
\+\ &{} &~~~~~~~~~~hep-lat/9711025.  \cr 

\+ [2&7] &E.V.~Shuryak and J.J.M.~Verbaarschot, {\it Phys.~Rev.} D52 (1995)
295. \cr

\+ [2&8] &SMC collaboration, {\it Phys.~Lett.} B412 (1997) 414. \cr  

\+ [2&9] &G.~Altarelli, R.D.~Ball, S.~Forte and G.~Ridolfi, {\it Nucl.~Phys.}
B496 (1997) 337.  \cr

\+ [3&0] &SMC collaboration, {\it Phys.~Rev.} D58 (1998) 112001. \cr 

\+ [3&1] &L.~Trentadue and G.~Veneziano, Phys.~Lett. B323 (1994) 201. \cr

\+ [3&2] &D.~de Florian {\it et al.}, Phys.~Lett. B389 (1996) 358. \cr 
 
\+ [3&3] &M.~Grazzini, L.~Trentadue and G.~Veneziano, {\it Nucl.~Phys.}
B519 (1998) 394. \cr

\+ [3&4] &A.~De Roeck and T.~Gehrmann (eds.), Proceedings of 1997 
Workshop,  \cr
\+\ &{} &~~~~~~~~~~{\it Physics with Polarized Protons at} HERA, 
hep-ph/9711358.\cr 
   
\vfill\eject

%}

\bye